\documentclass[sn-mathphys]{sn-jnl}

\jyear{2023}%

\theoremstyle{thmstyleone}%
\theoremstyle{thmstyletwo}%

\theoremstyle{thmstylethree}%

\begin{document}
\title[Rethinking negative sampling in content-based news recommendation]{Rethinking negative sampling in content-based news recommendation}


\author[1,2,3]{\fnm{Miguel Ângelo} \sur{Rebelo}}

\author*[4]{\fnm{João} \sur{Vinagre}}\email{joao.vinagre at ec.europa.eu}
\equalcont{Work partially conducted while author was affiliated with INESC TEC, Porto, Portugal and the University of Porto, Portugal.}
\author[5,6]{\fnm{Ivo} \sur{Pereira}}
\author[3,6]{\fnm{Álvaro} \sur{Figueira}}

\affil[1]{\orgname{Promptly Health}, \orgaddress{\city{Porto}, \country{Portugal}}}

\affil[2]{\orgdiv{Population Genetics \& Evolution}, \orgname{i3s}, \orgaddress{\city{Porto}, \country{Portugal}}}

\affil[3]{\orgdiv{Faculty of Sciences}, \orgname{University of Porto}, \orgaddress{\city{Porto}, \country{Portugal}}}

\affil[4]{\orgdiv{Joint Research Centre}, \orgname{European Commission}, \orgaddress{\city{Seville}, \country{Spain}}}

\affil[5]{\orgdiv{ISEP}, \orgname{Polytechnic of Porto}, \orgaddress{\city{Porto}, \country{Portugal}}}

\affil[6]{\orgname{INESC TEC}, \orgaddress{\city{Porto}, \country{Portugal}}}


\abstract{News recommender systems are hindered by the brief lifespan of articles, as they undergo rapid relevance decay. Recent studies have demonstrated the potential of content-based neural techniques in tackling this problem. However, these models often involve complex neural architectures and often lack consideration for negative examples. In this study, we posit that the careful sampling of negative examples has a big impact on the model's outcome. We devise a negative sampling technique that not only improves the accuracy of the model but also facilitates the decentralization of the recommendation system. The experimental results obtained using the MIND dataset demonstrate that the accuracy of the method under consideration can compete with that of State-of-the-Art models. The utilization of the sampling technique is essential in reducing model complexity and accelerating the training process, while maintaining a high level of accuracy.  Finally, we discuss how decentralized models can help improve privacy and scalability.}




\maketitle

\section{Introduction}

The large amount of available news sources and articles, coupled with 
very fast update cycles, creates an atmosphere of information overload that makes it hard for readers to keep track of news that are most relevant to them~\cite{bib2}. The power of personalized 
news retrieval can be extremely helpful in improving the users' overall satisfaction with the service. By carefully selecting items to users -- in this case, news articles -- Recommender Systems (RS) bring the most relevant items to the attention of users.

Recommender systems for news typically rely on user click data that consists of the positive interactions between users and news items. Because machine learning algorithms struggle to learn from positive data without a negative counterpart -- i.e. disliked or otherwise irrelevant items --, it is common practice to randomly sample negative examples to balance the learning data.
In this paper, a negative sampling technique is proposed, that fuels a Decentralized Neural News Recommendation system (DNNR) by providing better implicit negative examples for the model to train on and learn user patterns. News Recommender Systems (NRS) have certain characteristics related to their business model that are not often, or at all, observed in other domains. The key difference is the speed at which the relevance of the items decay. Unlike item recommendation in music, movies, or the retail market, for example, the relevance of news articles can change very rapidly concomitant with daily happenings and events \cite{karimiNews2018}. This leads to a permanent item \textit{cold-start} problem, since 
recent news items to recommend have few interactions. Fortunately, news are content-rich, and recent advances in natural language processing (NLP) provide excellent tools to extract rich and compact representations directly from natural text. 
These content-based representations can compensate for the scarcity of interactions of new items.

Another relevant aspect of our work consists of the decentralized nature of our proposed method. We have now arrived in an information-centric age, where computing power is unevenly distributed between provider infrastructure and user devices, where most data is generated \cite{DLedge}. Centralized computing power, where most computation involving the training of RS is done, need to efficiently manage and process these large quantities of data, produced in a widely distributed system, which raises some issues:

\begin{itemize}
    \item Cost: To train models and do inference on centralized computing power requires the transmission of massive amounts of data;
    \item Latency: the delay to access the provider's computing infrastructure power and storage is generally not guaranteed, and might restrain some solutions that are more time-critical. 
    \item Privacy: training models requires a lot of private information to be carried, raising privacy issues. Organizations with large amounts of user data heightens the risk of illegitimate data use or hazardous private data leaks. 
\end{itemize}

Under these circumstances, on-device or edge computing offers advantages by hosting some computation tasks close to the data sources and end users. When combined with centralized computing it can: alleviate backbone network, by handling key computation tasks without exchanging data with the central computing cluster; and allowing for more agile service response, by reducing or removing the delay of data transmissions \cite{DLedge}. It also has the potential to provide better privacy guarantees, while simultaneously granting users a finer control over processes involving their personal data.

Our approach is to train a different model for each user, which makes it trivial to decentralize. This has the potential to offload computation to the user realm, which besides reducing resource requirements from the provider's side, also improves privacy and user autonomy -- e.g. by allowing users to choose and fine-tune models to their needs. 

Summarizing, we our contributions are the following:
\begin{itemize}
    \item We introduce a personalized negative sampling technique for text-based recommendation, considerably improving model accuracy with respect to the state of the art;
    \item We propose a decentralized training strategy based on the idea of training a separate lightweight neural recommender for each user.
    \item We study the trade-offs between the negative sample size and predictive performance, as well as training and prediction times. 
\end{itemize}

We organized the paper as follows. In Section \ref{sec:related_work}, we provide a review of related work.

\section{Related Work}
\label{sec:related_work}

According to Jannach et al. \cite{RSlandscape2012}, Collaborative Filtering (CF) methods are the most common approach in the RS literature. This is explained by their domain-agnostic application and good overall performance without much information about the business model. Howbeit, things change when it comes to NRS. An analysis of 112 papers that propose one or more recommendation algorithms shows that 59 chose content-based approaches by creating reader's profiles based on past documents of interest and recommending articles that fit the user's pattern \cite{karimiNews2018}. This can be explained by the fact that the main content of news is text, which can be analysed to extract information. In addition, users and community features can also be used, although personal information should be avoided for ethical reasons. 
Furthermore, because reality changes constantly and people's preferences and interests vary over time, NRS have to keep the user profiles updated.

Akin to other domains, the information gathered from a user can be {\itshape explicit} preference information, such as a score in a rating scale, or {\itshape implicit}, by simply observing the user’s  behaviour, such as reading an article, sharing it, printing it, or commenting on it \cite{karimiNews2018}. 

As an example, The Athena news recommendation system \cite{IJntema} mostly relies on content information. The user profiles are constructed from a set of concepts from the articles the user has read, resulting in a vector with the distinct weighted concepts for applying distance metrics and semantic searches. A similar approach was used for the Ontology Based Similarity Model (OBSM) \cite{rao2013}, which calculates news-user similarity based through ontological structures, with user profiles having a \emph{bag-of-concepts} format with DBPedia\footnote{\url{https://www.dbpedia.org/}} as a knowledge base in the background.

Some approaches suggesting segmenting users according to their demographic information and article read patterns, weighted term vectors from the topics of the read articles \cite{demographic, ahn2007}. 

There are also models that solely rely on click behavior (interactions), like the ones that characterize the Google News Personalization system, which predicts the relevance of an article using both a long-term CF model and a short-term model based on article co-visitations \cite{googlenews2007}. More recently, an alternative approach was implemented, a Bayesian framework for predicting current news interests from the past predilections of each user and the community trend, combining content-based analysis for the construction of user profiles with an existing CF mechanism to generate personalized recommendations \cite{liu2010}.

Questions arise regarding if to or how to consider long-term and short-term preferences, for balancing the importance of each article view represents an important point of discussion in news recommendation \cite{vietnam}. There are questions whether two separate models should be built or else a time-decay factor should be included in an integrated model \cite{karimiNews2018}.

Lately, novel neural network designs have made considerable progress. Neural news recommendation model with personalized attention (NPA) \cite{NPA} is a news recommendation model with personalized attention \cite{attention}, that uses convolutional neural networks (CNN) to learn hidden representations of news articles based on their titles and learns user representations based on the representations created for their clicked articles. In addition, a word-level and a news-level personalized attention are used to capture different informativeness for different users. Deep knowledge-aware network (DKN) for News Recommendation is a deep learning model which incorporates information from a knowledge graph for better news recommendations \cite{DKN}. It applies knowledge graph representation learning and a CNN framework to combine entity embedding with word embedding and generate a final embedding vector for a news article. An attention-based neural scorer is used for click prediction. NRMS \cite{NRMS} is considered to be the state of the art, with its following variants \cite{wuPLM, negativefeedback}. It consists of a news encoder and a user encoder. It uses multi-head self-attention networks to learn news representations from titles, to model the interaction between words and applies the same principle to learn user representations, by capturing the relatedness between the news read by the user. In addition, additive attention is also used to learn more informative news and user representations by selecting important words and news. It has proven to be very effective. One of its biggest downsides is its black-box nature. 
The term {\itshape cold-start} refers to the situation where there is seldom to none information about user preferences or no information about a new item. 

For a {\itshape cold} user, one general approach is to incorporate additional information about the user's context. That can be the location, time of day or demography. An alternative to this is to incorporate features from the news articles for assessing their relevance to an hypothetical generic user, such as the freshness of the news article or its popularity. The YourNews system \cite{ahn2007}, as an example, starts by showing only recently published news, for the penalization process only starts after the first interaction. 

For a {\itshape cold} item, a content-based approach uses the data from the article to compare with the past content-wise preferences of the individual reader, which can solve the problem. So, the information contained in the articles can be extracted and analysed without it ever being read by anyone, which makes it instantly recommendable without the need for past interactions. This content can be, for example, {\itshape named entities} \cite{karimiNews2018}.

The belief that the user interest modeling still can be improved, without altering the model nor the embedding process, is one of the premises of this work. Many existing methods for training news RS solely rely on the implicit feedback from user clicks to infer their interests, interpreting the unclicked news as negative samples with a uniform probability -- the \emph{missing-at-random} assumption. In the past few years, some contributions have shown that this assumption rarely applies to real-world cases. However, preferences are far from a binary choice of either or, and there can be the case where every news presented to the user could be interesting, but the case was that he or she only picked one. It is also difficult to accurately sample negative examples without explicit user feedback. 
The incorporation of negative feedback inferred from the dwelling time of news reading was proposed in \cite{negativefeedback}, to distinguish positive and negative news clicks, via a combination of transformer and additive attention network. The use of factorization machines where also proposed to get negative samples from implicit feedback data when content information cannot be leveraged \cite{negsample}. This technique has also been used to reduce the amount of negative samples \cite{negsample2}. In \cite{Vinagre2015} negative items are sampled based on how far back in time they were interacted with. In this paper, we propose a new way to approach the negative sampling issue and tackle the training process for news interest modelling, 
by sampling negative examples that are naturally far away from the user's preferred items in the embedding space. 

For efficient news recommendation, text modeling is the key for understanding news content. Existing news recommendation methods usually model news texts based on traditional NLP \cite{NRembedding2017, DKN, finegrained, NRMS, NPA}. There are multiple examples of complex networks that use news representations, in the form of embeddings, from words and entities present in titles or the corpus of the news. However, these techniques do not capture the semantic relationships between words, which results in a shallow representation of the news content.

The introduction of pre-trained language models (PLM) revolutionized NLP, with great text modeling, performance and versatility. Usually, PLMs are pre-trained on a large unlabeled corpus via self-supervision to encode universal text information, and with the aid of their deeper networks, may have greater ability in modeling the complex contextual information in news text \cite{PLMsurvey}. 

Regarding decentralized recommendation models, -- see, for example, \cite{Canny2002,Han2004,Miller2004,Eichinger2019} -- the main focus is on learning collaborative filtering protocols for peer-to-peer networked communities. Algorithms are distributed and exclusively based on information exchanged between peers (users) -- {\itshape i.e.} without orchestration by a central entity. These proposals focus on collaborative models, that essentially exploit patterns in the user-item interactions. In this sense, our proposal is different, since we use a purely content-based approach. However, we borrow the idea of \emph{personal recommenders} proposed with PocketLens \cite{Miller2004}, since our motivations are very similar.

\section{Methods}
\subsection{Multi-network training framework}
\label{sec:methodology}

Since the main contributions of this work are two-fold -- the distributed model architecture and the negative sampling technique that supports it --, the general training framework is explained before its individual components, {\itshape i.e.} the neural network design, the embedding process using PLMs and the negative sampling technique. These components combined create the foundations for the proposed training framework, which changes the way these processes can be approached to tackle issues related to scalability, latency, cost, and most importantly, privacy.

\begin{figure}[ht]
  \centering
  \includegraphics[width=\linewidth]{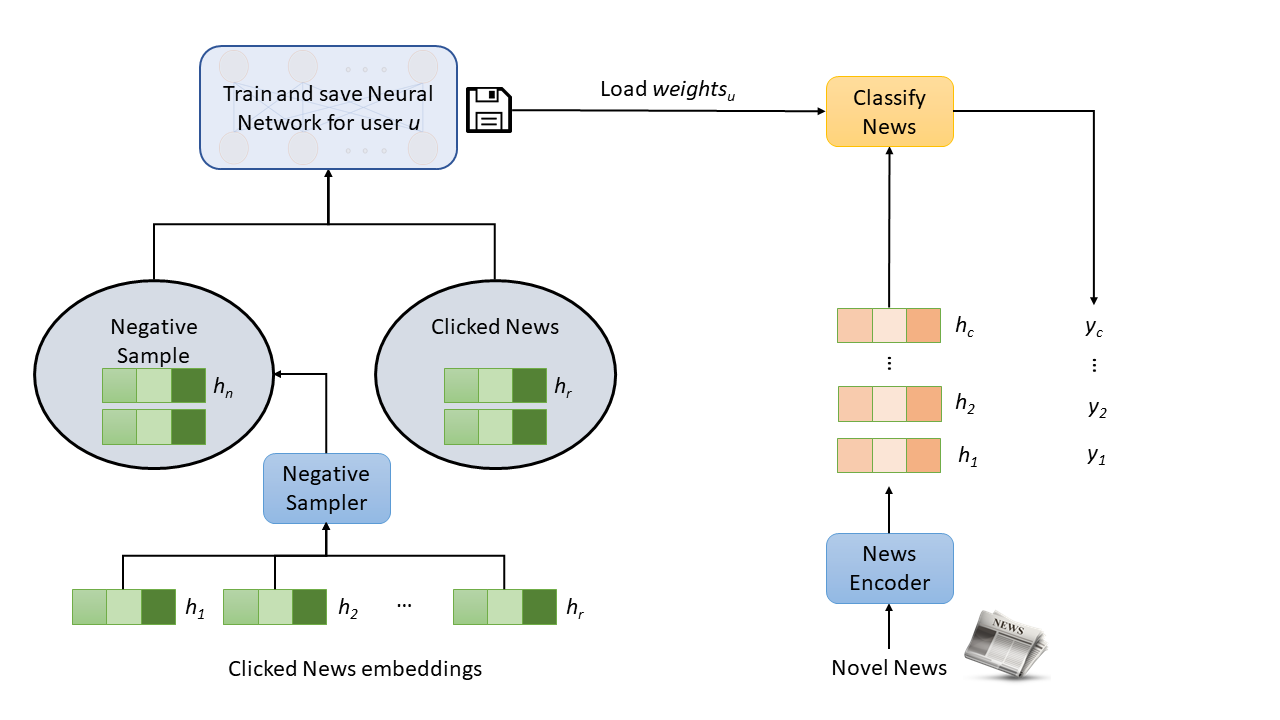}
  \caption{Training with negative samples Framework}
  \label{fig:TFW}
\end{figure}

\subsection{News Recommendation Training Framework}
The main components in this training framework include a news encoder to transform news variables into fixed-size tensor embeddings, a negative sampler to generate synthetic negative samples based on the user's history and the classification module. The news encoder processes \textit{r} read (or clicked) news for each user and creates its embeddings denoted as \begin{math}
  [h_1, h_2, ..., h_r]
\end{math}. The news encoder also creates embeddings for the \textit{c} candidate news: 
\begin{math}
  [h_1, h_2, ..., h_c]
\end{math}. For each user \textit{u}, the negative sampler creates a synthetic negative sample of size \textit{n}
\begin{math}
  [h_1, h_2, ..., h_n]
\end{math}
based on the \textit{r} read news, with 
\begin{math}
  n = r
\end{math} to achieve a balanced sampling every time. A small neural network for each user is then trained with the \textit{Synthetic Pool} (the users' history coupled with the synthetic negative feedback - explained in more detail in subsection~\ref{subsection:ns}) and saved for later, when novel candidate news will need to be scored. Figure~\ref{fig:TFW} illustrates the training framework with all of its components.

\subsection{Neural-network design}


To achieve fast training and prediction in a decentralized setting, neural networks need to be lightweight. Network layers were kept to 10 for quick \textit{epochs}. From the total number of n-features, the first dimensions corresponding to the PLM-embedded titles run through 4 initial layers to reduce its dimensionality from 384 to 64. Then, this 64-dimensional vector is concatenated to the rest of the original vector to continue through the feed-forward network. All layers have rectified linear unit (ReLU) activation functions, excepting the forth layer, which uses a Hyperbolic Tangent (Tanh) activation function, before the concatenation. 

\begin{figure}[ht]
  \centering
  \includegraphics[width=340pt]{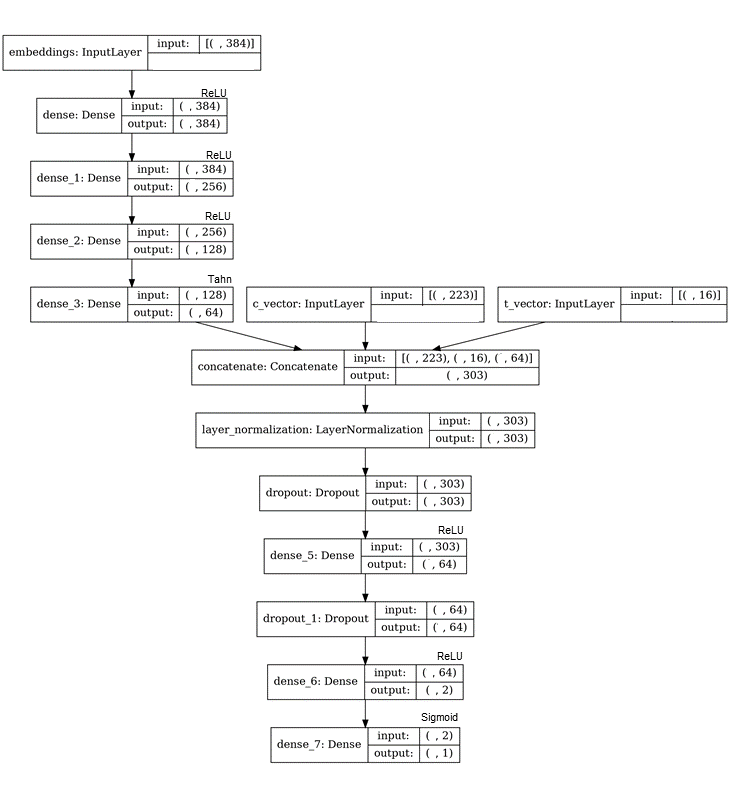}
  \caption{Feed-forward Neural Network Architecture}
  \label{fig:ffnn}
\end{figure}

After concatenation, the activations of the previous layer for each given example are normalized by passing through a normalization layer. Two Dropout points are added, randomly zeroing 0.2 of the elements of the input tensor, which has proven to be an effective technique for regularization and preventing the co-adaptation of neurons as described in \cite{dropout}. It outputs to a 2-dimensional vector that can be interpreted as the inverse-sigmoid of the threshold. Hence, a sigmoid function is applied to the output for class 1, which corresponds to the \emph{read} class, to get the read probability $\textit{Read Probability}(output[,1]) = Sigmoid(x) = \frac{1}{1+exp(-x)}$. This transforms the problem into a regression, which is important to order the recommendations based on the clicking probability. That is why the loss function used is mean squared error (mse) instead of binary cross-entropy. Figure~\ref{fig:ffnn} displays a graphic representation of these small neural networks.

\subsection{PLM-powered Embeddings}

\begin{figure}[ht]
  \centering
  \includegraphics[width=400pt]{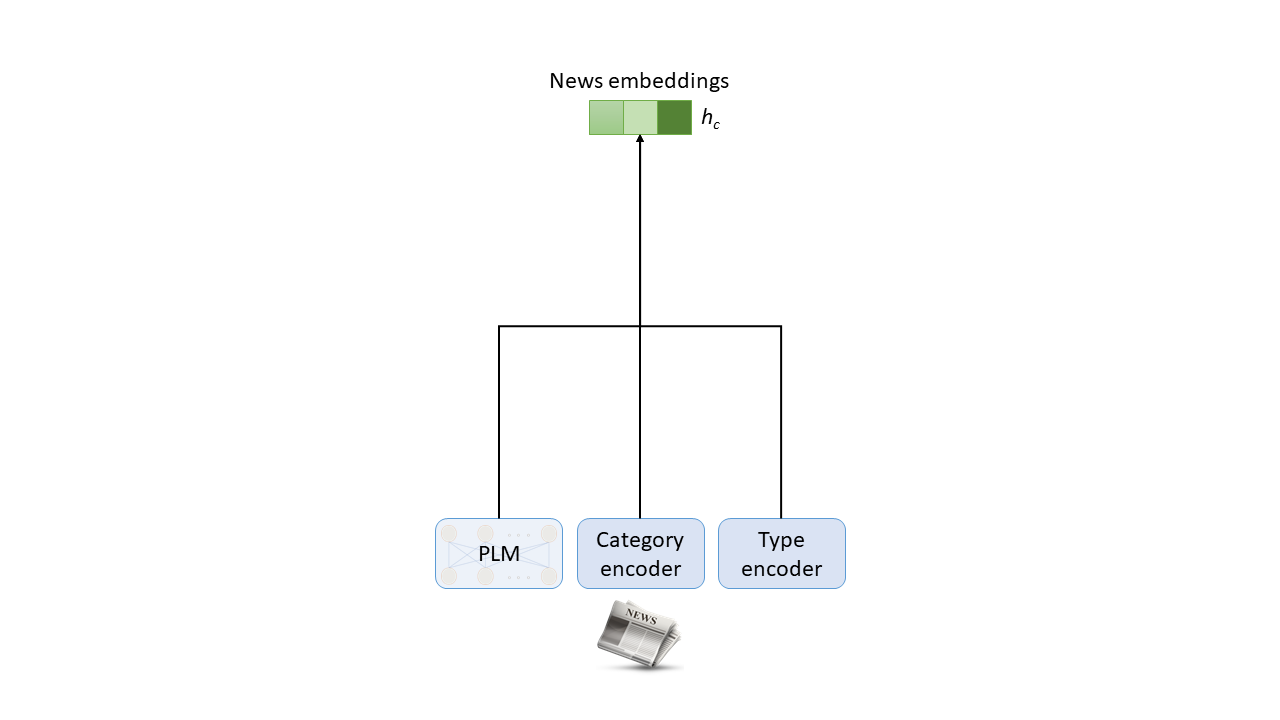}
  \caption{PLM powered News encoder}
  \label{fig:PLM} 
\end{figure}

Pre-trained Language Models (PLM) were used to empower the content embedding process. In this case, a deep self-attention distillation of a multilingual pre-trained model \cite{miniLM} was used due to its speed to size relationship. Figure~\ref{fig:PLM} illustrates the news encoder module, which embeds news titles to a 384-dimensional vector using a PLM, and one-hot encodes the news category and type into fixed-size vectors. These embeddings have a crucial role in the whole process:
\begin{itemize}
    \item they are required for the negative sampling method proposed.
    \item the three fixed-size vectors are concatenated and fed to the neural networks to train.
\end{itemize}

\subsection{Negative Sampling Technique and Model Training}
\label{subsection:ns}
The creation of a synthetic negative sample for each user was approached as essential to the success of the proposed distributed approach, using the history for each user as the reference. To achieve this, the news title embeddings are indexed to search in the L2 space the farthest news from the reference centroid of the user. The reference centroid is computed as the averaged embeddings of all titles from the user's history. The squared Euclidean (L2) distance is monotonic as the Euclidean distance, but if exact distances are needed, an additional square root of the result is needed. The inner product was used for maximum inner product search. It is not by itself cosine similarity, unless the vectors are normalized (lie on the surface of a unit hypersphere). 
For $\textit{l}^2$-normalized vectors $x,y$,
$||x||_2=||y||_2=1$, we have that the squared Euclidean distance is proportional to the cosine distance (equation~\ref{equation:1}).
\begin{equation}
\label{equation:1}
||x-y||_2^2=(x-y)^T(x-y)
=x^Tx-2x^Ty+y^Ty
=2-2x^Ty
=2-2cos(x,y)
\end{equation}

\begin{figure}[ht]
  \centering
  \includegraphics[width=\linewidth]{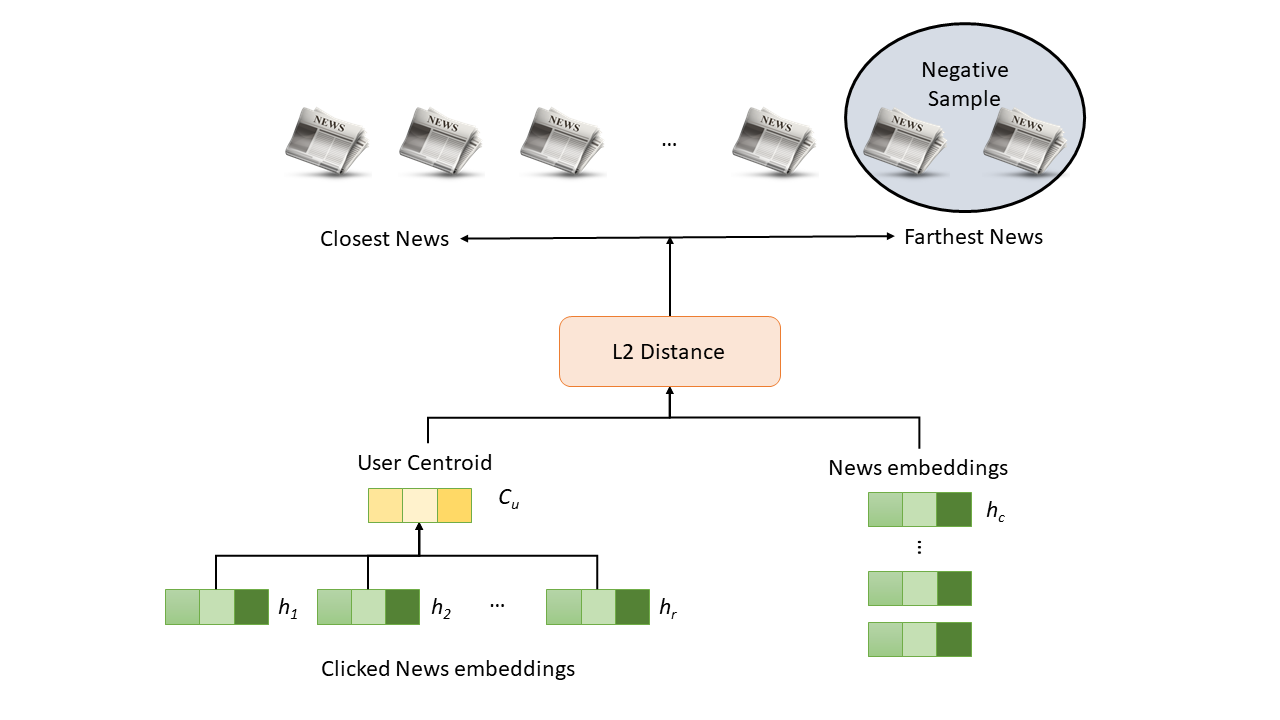}
  \caption{Synthetic Negative sampling method}
  \label{fig:NS}
\end{figure}

Then we take the positive sample, which is simply the users' history and feed the pooled sample with positive and synthetic negative feedback to the model. We refer to this pooled sample as \emph{Synthetic Pool}. For a perfectly balanced dataset, the length of the negative sample matches the length of the positive one, which facilitates the learning process. A limit of $60$ more recent samples was introduced, since it was enough to capture user profiles while cutting on training time. The process can be condensed in a few lines (algorithm~\ref{algo1}), and by its graphical illustration, in figure~\ref{fig:NS}. 
To obtain a reliable evaluation of this technique's performance, it was compared against two other variants: (i) a model trained only on news impressions -- which contain news that were presented to the user, as well as an indicator to whether the user clicked or not \cite{datasetMIND} --, taking the non-clicked news as negative feedback; and (ii) a model trained on the data with random news taken from the pool of unread news as negative feedback. In other words, the comparison was made by testing the models trained on three different samples - Synthetic Pools, news impressions, and random sampling - against the news impressions.

\makeatletter
\def\BState{\State\hskip-\ALG@thistlm}
\makeatother

\begin{algorithm}
	\caption{Create Synthetic Negative Samples}\label{algo1}
	\begin{algorithmic}[1]
		\Procedure{algorithm}{}
		\BState $\textit{embeddings} \gets \text{embeddings for news read by}\textit{users}$
		\BState $\textit{embeddings} = \text{L2-norm(embeddings)}$
		
		\BState \emph{For each user u}:
		\State {$\textit{embedding vectors for user u} \gets \textit{embeddings}(u)$}
		
		\State {$\textit{User centroid} = \overline{\textit{embedding vectors for user u}}$}

		\State {$ \textit{inner product indices} \gets \textit{Sort(} \langle centroid \; , \; embeddings \rangle \textit{)}[indices]$}
		
		\State {$\textit{Sample Length}\gets \textit{Length(embedding vectors for user u)}$}
		
		\State {$\textit{Synthetic Pool for user u}\gets \textit{inner product indices[Sample Length]}$}
		
		\State \Return \textit{Synthetic Pool for user u}
		\EndProcedure
	\end{algorithmic}
\end{algorithm}

\section{Experiments and results}
\label{sec:experiments}

The experiments were conducted on a real-world dataset from Microsoft, the famous MIND dataset \cite{datasetMIND}. This dataset is mono-lingual (english) and has data from a news aggregator -- {\itshape i.e.} it includes multiple news sources --, having high content diversity. The MIND-small dataset has anonymized behavior logs from the Microsoft News website. It contains click histories and impressions logs of 50,000 randomly sampled users who had at least 5 news clicks during 6 weeks from October 12 to November 22, 2019. Table~\ref{tab:dat} contains the MIND dataset statistics. When it comes to the content of this dataset, table~\ref{tab:content} shows an example line from the behaviors file, from which the history and impressions for each user were extracted. 

\begin{table}[ht]
\centering
  \caption{MIND Dataset Statistics}
  \label{tab:dat}
  \begin{tabular}{ccll}
    \toprule
    Parameters & Value\\
    \midrule
    \# Users & 50,000\\
    \# News & 51,282\\
    \# Unique Interactions & 926,058\\
    \# Impressions & 1,804,520\\
  \bottomrule
\end{tabular}
\end{table}

\begin{table}[ht]
\centering
  \caption{MIND Dataset Example Content}
  \label{tab:content}
  \begin{tabular}{ccll}
    \toprule
    Column & Content\\
    \midrule
    \ Impression ID & 91\\
    \ User ID & U397059\\
    \ Time & 11/15/2019 10:22:32 AM\\
    \ History & N106403 N71977 N97080 N102132 N97212 N121652\\
    \ Impressions & N129416-0 N26703-1 N120089-1 N53018-0 N89764-0 N91737-0 N29160-0\\
  \bottomrule
\end{tabular}
\end{table}

\begin{figure}[ht]
  \centering
  \includegraphics[width=220pt]{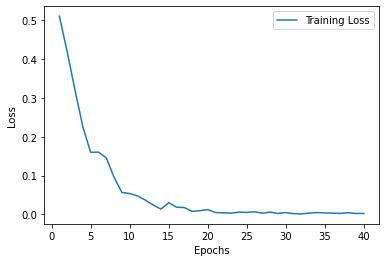}
  \caption{Loss per epoch during FFNN training}
  \label{fig:losses}
\end{figure}

Due to a limit of GPU memory, batch sizes were kept to 64 (60 in the case of DNNR) which is size used in \cite{NRMS}. The number of epochs per user were set to 15, since loss values stabilize at around 15 iterations (Figure~\ref{fig:losses}). 

\subsection{Getting in touch with the data}

\begin{figure}[ht]
  \centering
  \includegraphics[width=\linewidth]{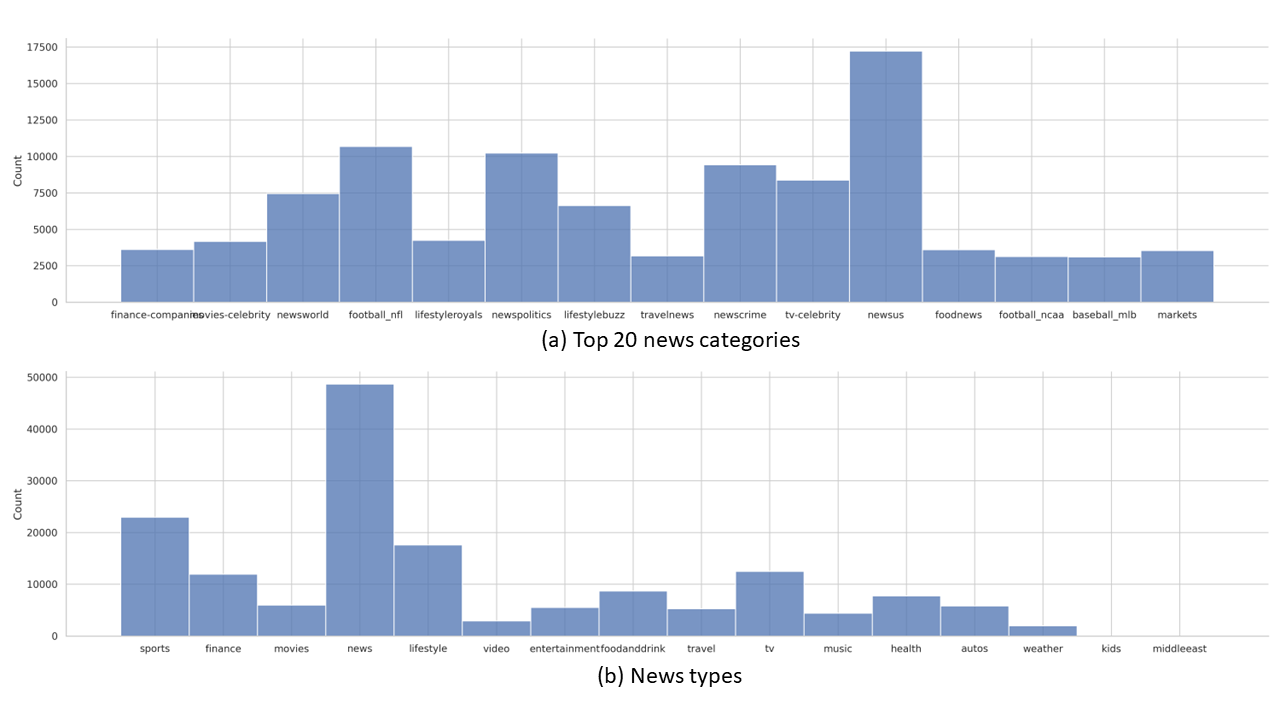}
  \caption{(a) Top 15 read news categories from the MIND dataset. (b) News types from the MIND dataset}
  \label{fig:bartypes}
\end{figure}

To  get to know the dataset at hand, firstly we looked at the amount of different news types there were and the amount of different news categories. There are $16$ news types that accommodate $212$ news categories. This diversity is great, providing a good starting point for the construction of the feature vector. Figure~\ref{fig:bartypes} shows the barplot for the top 15 read news categories from the MIND dataset (a) and the news types from the MIND dataset (b). The most read news category is the 'newsus', News related to the United States, followed by NFL football, politics and crime, the expected subjects. As for the types, the predominant one is 'news', followed by sports and lifestyle. Although the distribution across different categories and types is not balanced (as expected), every category (in types and categories) are fairly well represented (except for the types 'kids' and 'middleeast', that apparently don't get that much traffic. 

User activity was further characterized with some descriptive statistics applied to the amount of items each user read, the amount of different types of news consumed and the amount of different news categories (Table~\ref{tab:data_table}). 

\begin{table}[ht]
\centering
\caption{Descriptive statistics on user activity.}
\label{tab:data_table}
\begin{tabular}{@{}llll@{}}
\toprule
Metric & Items read & Types read & Categories read \\ \midrule
mean   & 28.10      & 6.97       & 13.68           \\
std    & 30.26      & 3.23       & 9.53            \\
min    & 1          & 1          & 1               \\
25\%   & 8          & 4          & 6               \\
50\%   & 19         & 7          & 12              \\
75\%   & 37         & 9          & 19              \\
max    & 343        & 14         & 73              \\ \bottomrule
\end{tabular}
\end{table}

\begin{figure}[ht]
  \centering
  \includegraphics[width=200pt]{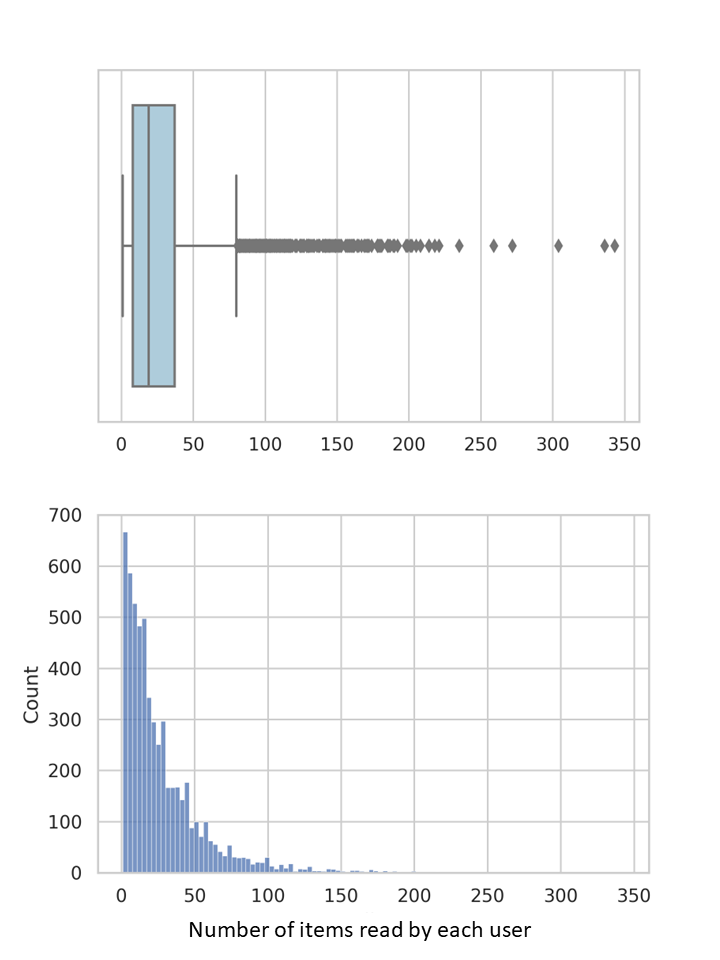}
  \caption{Distribution analysis of the number of items read by each user}
  \label{fig:data_items}
\end{figure}

It is possible to observe that most users read just $19$ news during the specified time-frame, which is not a lot. The distribution is fairly left-skewed (Figure~\ref{fig:data_items}), with users that read more than 80 news items ($37+1.5*(37-8)$) being considered outliers. The most avid reader in this sample reached $343$ readings, which is more history than we will ever need. However, there are many users with low item counts, which is not ideal to model their preferences. 

\begin{figure}[ht]
  \centering
  \includegraphics[width=200pt]{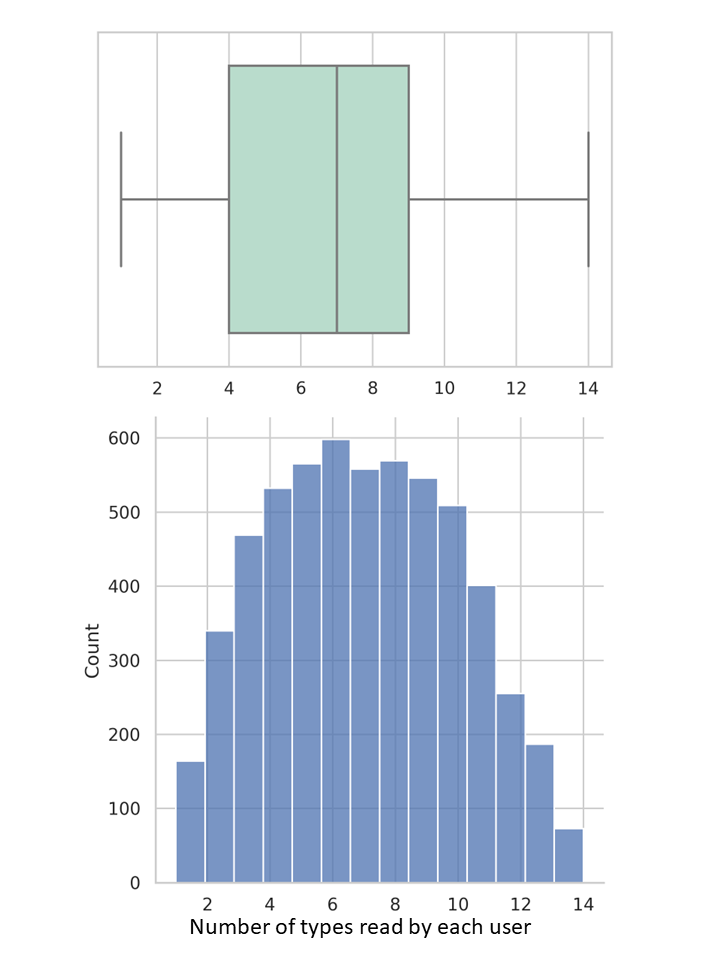}
  \caption{Distribution analysis of the number of different types of news read by each user}
  \label{fig:data_types}
\end{figure}

News articles are grouped within $16$ news types, which is good because it gives us another variable to infer user preferences other than just the plain text contained in the title. Although there are $16$ news types to chose from, the max number of different news types a client consumed were $14$ with most of them reading around $7$. The distribution (Figure~\ref{fig:data_types}) is close to symmetrical, with most of the clients reading less than half of the available news types, which is compatible with the varying preferences among clients.

\begin{figure}[ht]
  \centering
  \includegraphics[width=200pt]{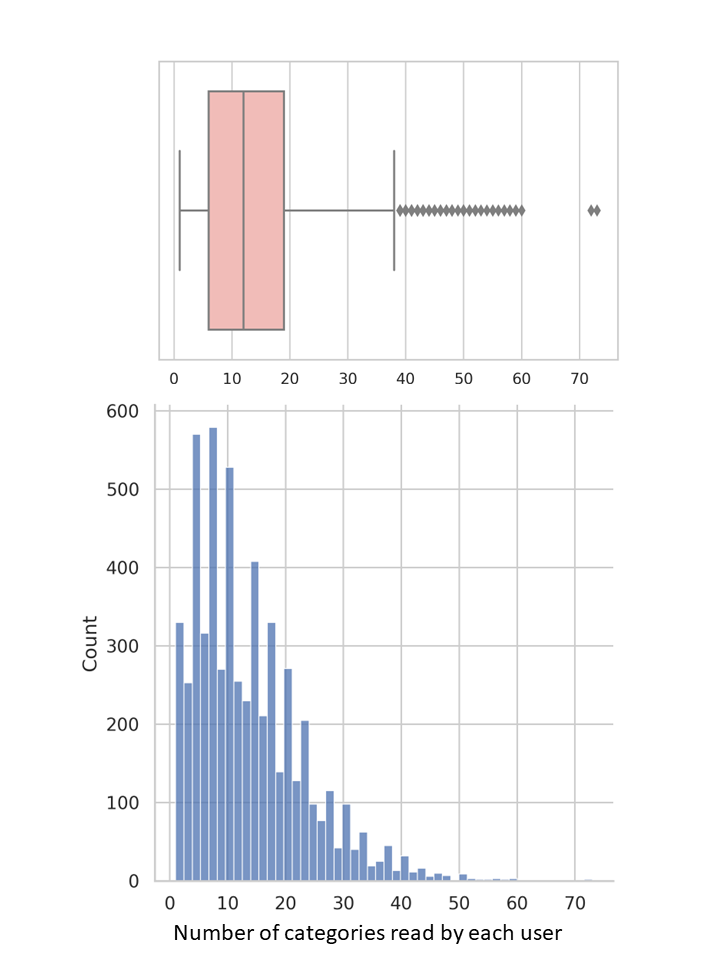}
  \caption{Distribution analysis of the number of different categories of news read by each user}
  \label{fig:data_category}
\end{figure}

Finally, news articles are distributed across $212$ different categories, which is a fairly good amount of dimensions to use when modelling reading preferences (Figure~\ref{fig:data_category}). We can see that most users spread around 12 categories, but the distribution is left-skewed, which means that there are some outliers which read across a much more diverse set of news categories, reaching a maximum of 73, with several outliers reading above 38 ($19 + 1.5 * (19 - 6)$). 

\subsection{Choosing the optimal number of max samples per user}

\begin{figure}[ht]
  \centering
  \includegraphics[width=\linewidth]{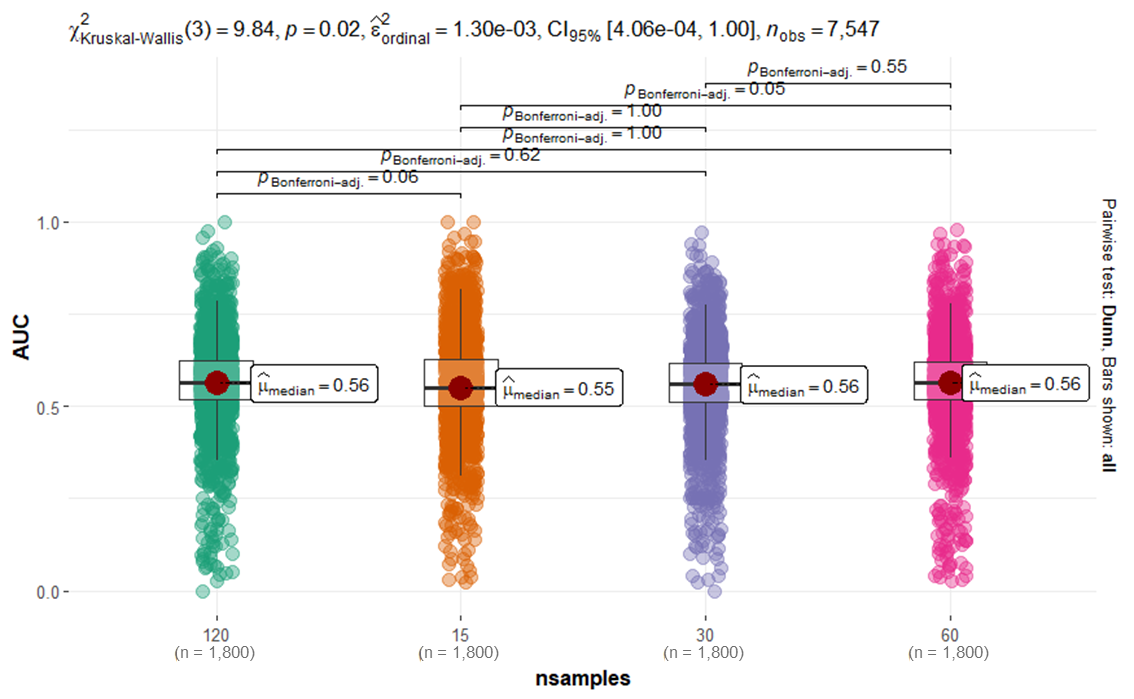}
  \caption{Max sample size effect on individual AUC}
  \label{fig:sampling_indAUC}
\end{figure}

It was already mentioned the decision to take the most recent samples (page views) up to 60 (see Section~\ref{subsection:ns}), if ever the user did read that much. However, here lies the explanation for that decision. Although it could be anchored just based on the analysis of the interaction between computational time and performance (area under the receiver operating curve - AUC) alone, picking the subjectively better sample size based on the trade-off between these two metrics, the decision would be better off if the differences in individual AUC between max sample sizes were statistically significant. Figure~\ref{fig:sampling_indAUC} sums up the Kruskal-Wallis test for different medians coupled with the Dunn's pairwise test between samples for 1800 samples. Due to the size of the samples, the visual differences are hard to pinpoint, but the $p_{Bonferroni-adjusted}=0.02$ indicates that at least one sample has a different median from the rest, it can be said that the four max sample sizes produce different effects. 

\begin{figure}[ht]
  \centering
  \includegraphics[width=\linewidth]{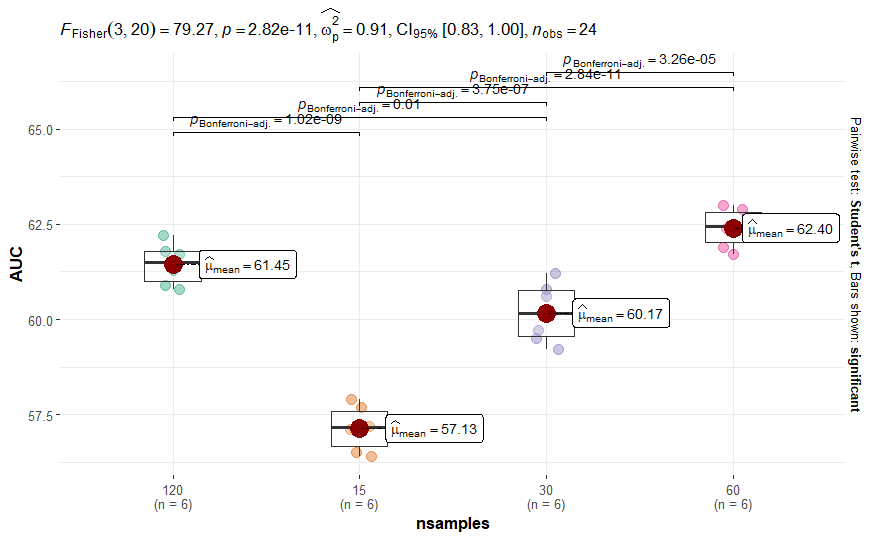}
  \caption{Max sample size effect on group AUC}
  \label{fig:sampling_groupAUC}
\end{figure}

From the Dunn's test the only statistically significant differences in median values (Bonferroni corrected) are between 120 and 15 max samples and between 15 and 60 samples, with 60 max samples achieving the highest individual AUC median, with no statistically significant difference detected between 60 and 120 max samples. Although the distributions appear to be relatively symmetric with a similar variance, here, due to the amount of samples (which up-eases the argument of statistical power), what interests the most is the median. That is why the Kruskal-Wallis test was chosen. To help decide the better threshold (between 30 and 60), an ANOVA test was performed to evaluate the difference between the group AUC (figure~\ref{fig:sampling_indAUC}). For this, six sample runs were performed for each threshold. Figure~\ref{fig:sampling_groupAUC} shows the difference in group AUC, and there is at least one mean that is significantly different from the rest. Here the one-way ANOVA seems to be the correct approach, since there are few samples (the mean is more interesting) and the distributions are approximately symmetric and homocedastic. Student's \textit{t} was performed for pairwise post-hoc testing. All differences are statistically significant except for the one between $15$ and $30$ ($p_{Bonferroni-adjusted}=0.10$). Most importantly, the clear choice seems to be $60$ samples, since it has a statistically higher group AUC when compared to both $30$ ($p_{Bonferroni-adjusted}=3.26e-05$) and higher $120$ although not statistically significant. 

Finally, it was inspected if the number of training samples, within a certain cap (tested for $60$ max positive samples per user) had a significant effect on the final AUC score for each user. For that, a simple linear model with a random sample of $500$ AUC values and the length of the training sample (s\_length) for each user was fitted and the coefficients analysed. Though the $p-value$ appears to be significant (see Table~\ref{tab:linreg}), the estimate for the sample length indicates a very small negative effect $-8.449e-06$ (more training samples are associated with lower AUC scores). So, if for each additional sample we can expect a decrease of this magnitude in the AUC score for a particular reader, even if a reader were to have $100$ more training samples, we should only expect to see a drop in the AUC score of approximately $0.85\%$. 
This does not change the fact that the result above stated is counter intuitive. We would have thought that if we had more information about a particular user, we could serve better recommendations and not worse. But avid readers tend to read across a more diverse spectrum of topics, and this diversity could be inserting more uncertainty in the modelling process. This being said, the possibility that readers with higher training sample counts might have bellow average performance, and thus significantly worse recommendations (leading to a worse user experience), is excluded.

\begin{table}[ht]
\centering
\caption{Linear Regression $AUC(s\_length_i) = \beta_0 + \beta_1(s\_length_i)$.}
\label{tab:linreg}
\begin{tabular}{@{}lllll@{}}
\toprule
            & Estimate   & Standar Error & t value & \textit{p-value (\textgreater{}|t|)} \\ \midrule
(Intercept) & 5.879e-01  & 1.169e-02     & 50.28   & \textless{}2e-16 ***                 \\
s\_length         & -8.449e-06 & 3.985e-06     & -2.12   & 0.0365 *                             \\ \bottomrule
\end{tabular}
\end{table}

\subsection{Impact of the Synthetic Negative Sampling}

To better understand the impact of this sampling method, the comparison was made between training the small user networks on three different types of samples and testing them on impressions. 

\begin{figure}[ht]
  \centering
  \includegraphics[width=\linewidth]{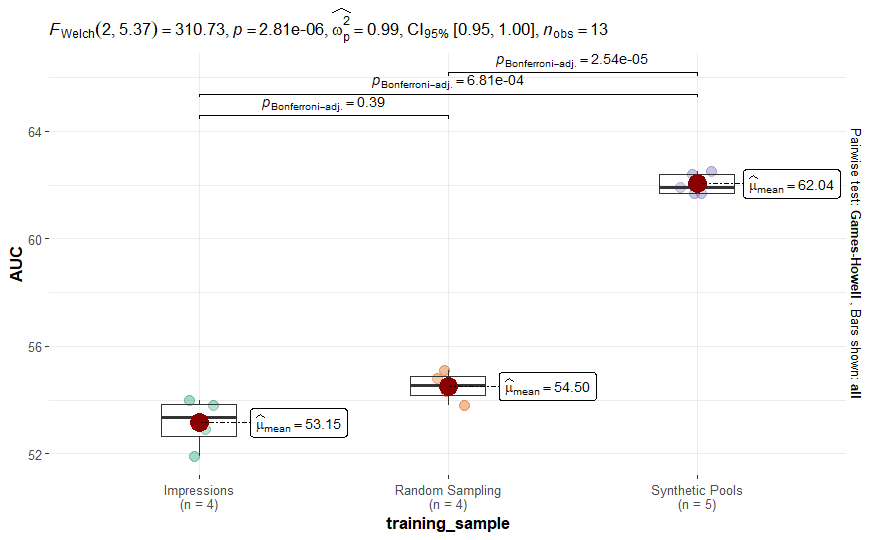}
  \caption{Influence of the Synthetic Pool: Group AUC comparison}
  \label{fig:influence}
\end{figure}

Figure~\ref{fig:influence} shows the difference in AUC when training the models on three different datasets: Impressions, Random Sampling, Synthetic Pools. While the simple neural networks trained on the news impressions cannot hold well, training these networks on the Synthetic Pools manages to model news profiles well enough to perform significantly better. Since the distributions are approximately symmetrical but do not have equal variances, the Welsh test was applied to check if the three population means are equal. Here, we can say that they are clearly not ($p=2.81e-06$), so at least one mean is statistically different than the others. Because group variances are unequal, the post-hoc pairwise test applied was the Games-Howell test, as an alternative to Tukey-Kramer. Although the difference between Impressions and Random Sampling is not statistically significant ($p_{Bonferroni-adjusted}=0.39$), there is a statistical difference between Synthetic Pools and Impressions or Random Sampling ($p_{Bonferroni-adjusted}=6.81e-04$, $p_{Bonferroni-adjusted}=2.54e-05$). 

\begin{figure}[ht]
  \centering
  \includegraphics[width=\linewidth]{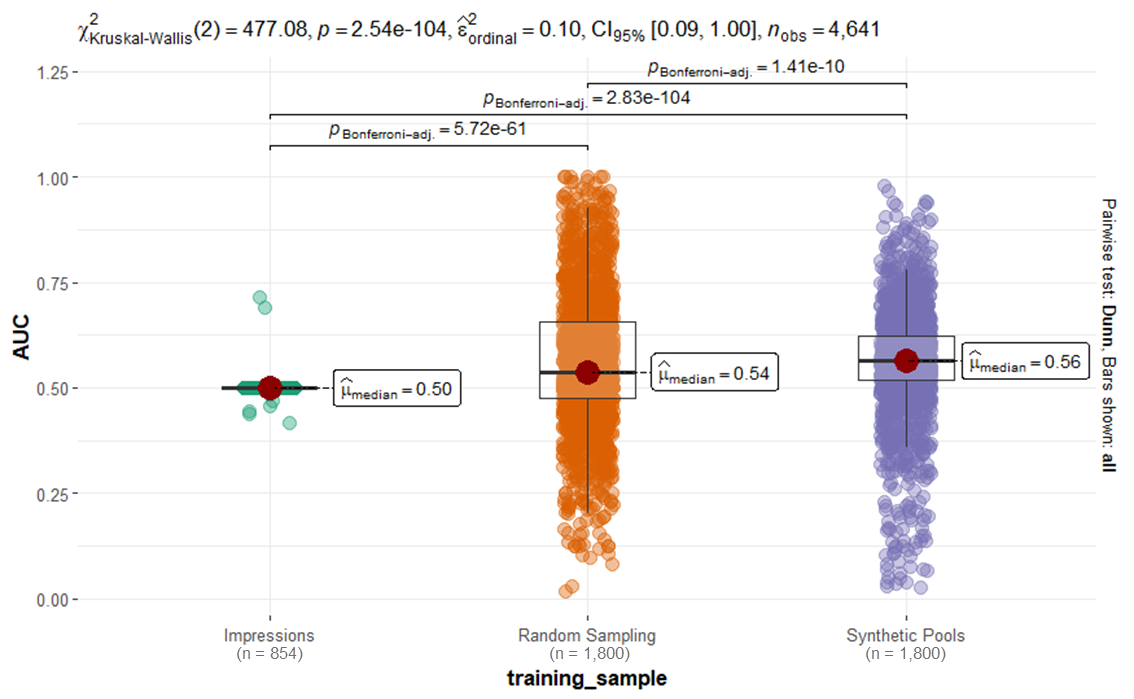}
  \caption{Influence of the Synthetic Pool: Individual AUC comparison}
  \label{fig:influence2}
\end{figure}

To have a clearer picture of what is going on, the individual AUC was also tested. Figure~\ref{fig:influence2} shows the difference in individual AUC for 1800 readers, when training the models on three different datasets (some cases failed using Impressions due to the inability to accurately model preferences, failing to identify even a single true positive). While for the group AUC the difference between Impressions and Random Sampling was not statistically significant, in the individual AUC there is a marked difference. Using the impressions with this method has terrible results, essentially making the prediction random, with a medium AUC of $50\%$. Using the Random Sampling to obtain negative samples improves a lot, with a statistically significant difference, but we can visually see that the variance is quite high. The Synthetic Pools seem to have two major consequences: significantly raise the individual AUC when compared to the Random Sampling ($p_{Bonferroni-adjusted}=1.41e-10$), and shortening the variance, which is desired since the results are more consistent across readers with different reading habits and routines.  

These results substantiate this technique's ability to efficiently model user preferences, provided that the Synthetic Pools assure a clear distinction between read news and putatively uninteresting news (possible by the synthetic negative feedback sampling method). It is noticeable that even a random sampling of negative samples from the pool of unread news can improve the training of these networks, achieving better mean performance than the networks trained on news impressions alone. 

\begin{figure}[ht]
  \centering
  \includegraphics[width=\linewidth]{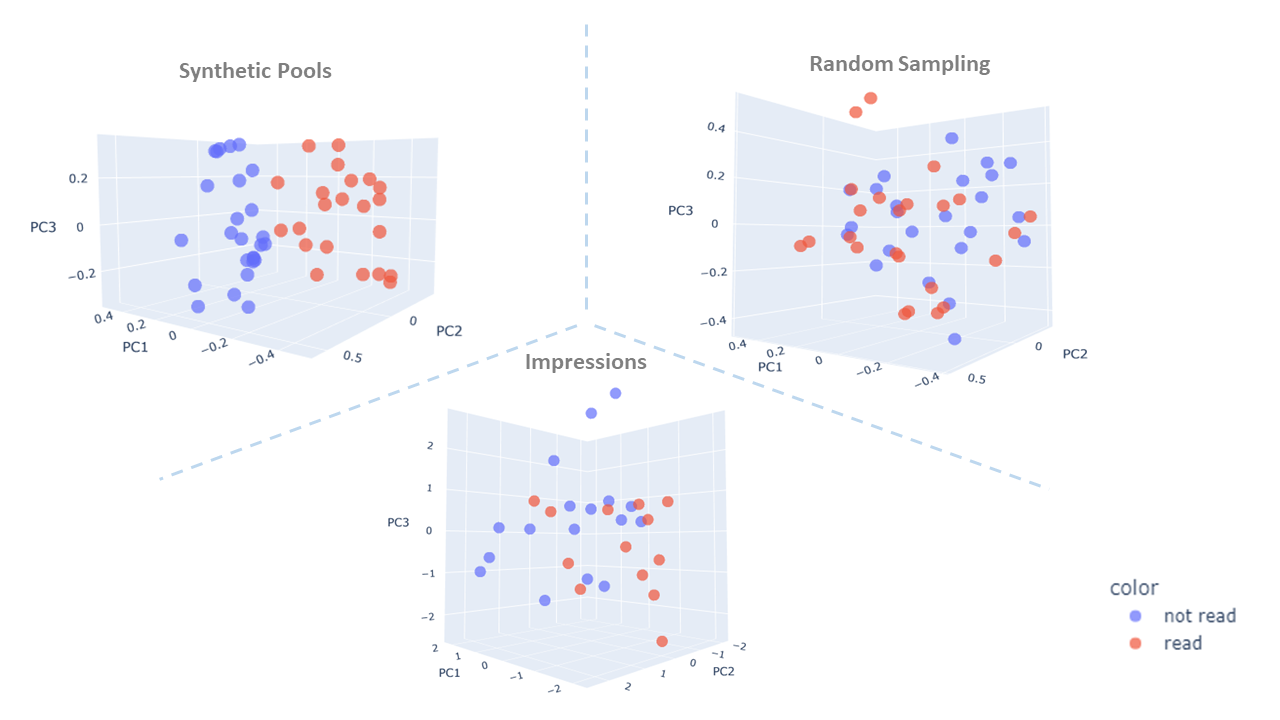}
  \caption{PCA of the Synthetic Pool VS Random negative sampling VS Impressions}
  \label{fig:sampling}
\end{figure}

Since the synthetic samples are gathered based on the dissimilarity between them and the user's historical news, it is easier to get better results when testing on this set. 
Conversely, the news impression set may not have so much dissimilarity between them, since they are based already on some kind of recommendation system, making it a harder set to capture user profiles and testing predictions. 
News impressions are collected from a pool of news that were presented to the user, with the information if the user clicked or not, it can already be filtered in a way that the user might have done a coin toss between two of them. So, some news marked as $0$ (not clicked) might have been of some interest to the user, but for some reason other than disliking the subject they were not clicked. 
By checking the first three principal components from the PCA analysis (figure~\ref{fig:sampling}), it is clear that we cannot observe clusters in the impressions, nor the random sampling sets. As for the Synthetic Pools, there is a marked distinction between the clicked historical news for an user, and its synthetic negative feedback. Thereupon, if we consider a classification task, it is expected that the latter sample would need a simpler model to attain the same accuracy, when compared with the other samples. This was the rationale behind the synthetic negative sampling technique. 

These results expose how data quality imposes limits on ML models, since the only thing that changes is the characteristics of the training sets.

\subsection{Offline Performance Evaluation}

\begin{figure}[ht]
  \centering
  \includegraphics[width=\linewidth]{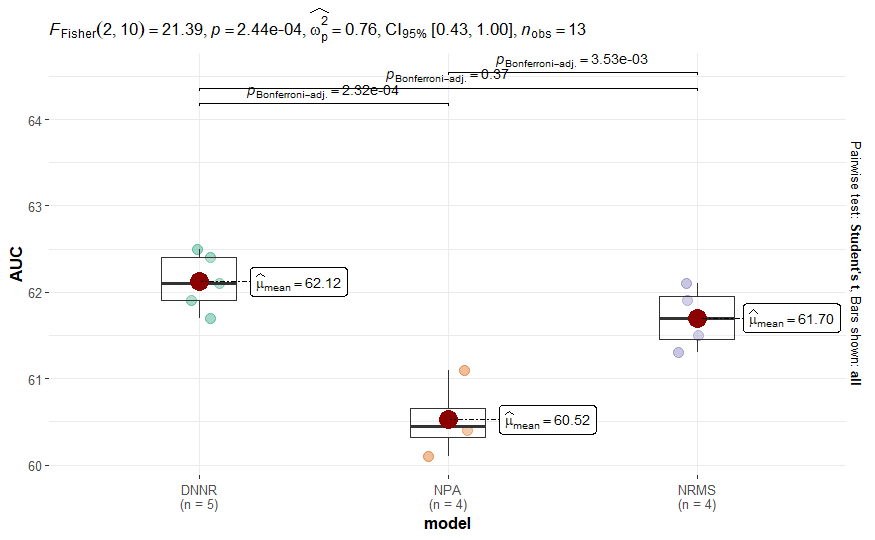}
  \caption{Comparison of AUC scores between the proposed approach against state of the art Models}
  \label{fig:perf}
\end{figure}

To measure the effectiveness of the proposed approach, the small networks were trained on the Synthetic Pools and tested on samples of impressions. This was compared to some of the most recent state of the art news recommendation methods, the NRMS \cite{NRMS}, the DKN \cite{DKN}, and the NPA \cite{NPA}. As seen before, the Synthetic Pools have a great impact when modelling small lightweight networks to capture user reading profiles, but the next question is whether or not it can compete with recent state of the art models. Figure~\ref{fig:perf} contains the mean values for AUC for each approach \cite{auc}. The results show that this DNNR approach can rival the best models to date, achieving a very similar performance.


These results are explained by the singular characteristics of the proposed approach. First of all, PLMs are stronger than the shallow models learned from scratch at text modeling, capturing semantic relationships \cite{wuPLM, bert}. We've chosen a multilingual PLM, the Multilingual MiniLM, which uses the same tokenizer as XLM-R but the transformer architecture is the same as BERT, which maps sentences to a 384 dimensional dense vector space for tasks like clustering or semantic search \cite{miniLM, bert2}. This was chosen due to having one of the best performance-speed relationship, and because multilingual models tend to outperform monolingual ones, which is a major advantage if we want to jointly train models to serve users in different languages \cite{wuPLM}. 
Second, since each user has its own network, all of the optimized weights and biases in the network were trained only on data directly related to a single user. 
Finally, the Synthetic Pools used to train the small individual networks make it easy to model each user's profile, requiring less from the model's architecture, since the quality of the data that is fed into the model is better (see figure~\ref{fig:sampling}). 

\subsection{Speed-Accuracy trade-off}
One key aspect to think about when designing a high throughput system that can infer on large amounts of data within a realistic amount of time is the computational time and its trade-off relationship with accuracy (here the word 'accuracy' is being used in the broader spectrum of the term). Although the design of the system allows for more clever implementations to achieve better computational times (see section~\ref{sec:discussion}), it should also be able to predict in bulk efficiently, if needed. As such, we need to consider the impact of the number of max positive samples per client on the computational times and the final AUC scores. As for the computational times, to get a more precise look at what is happening, we present the time it took to generate the Synthetic Pools (Synthetic Pooling time) and the time it took to generate the predictions. 

\begin{figure}[ht]
  \centering
  \includegraphics[width=\linewidth]{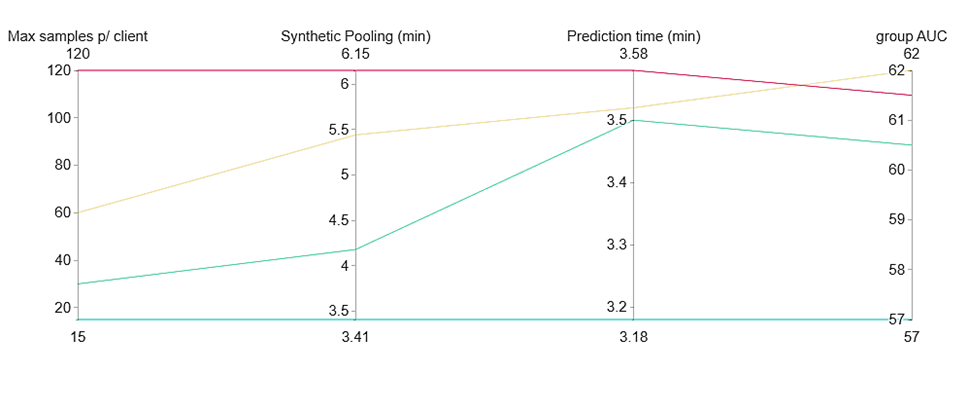}
  \caption{Comparison of AUC scores resulting from different max number of samples considered per client, and the corresponding Synthetic Pooling and prediction times (in minutes per 4000 clients). Starting with the max sample size per reader, this shows its effect on the time it takes to compute the Synthetic Pools as well as the toll it takes in the time needed to predict, and finally its effect on group AUC. Each max sample size cap has its own color.}
  \label{fig:time}
\end{figure}

Figure~\ref{fig:time} showcases these metrics for four different max sample sizes - $15$, $30$, $60$ and $120$ - and the resulting times and group AUC. The line at the bottom is the worst contender, $15$ max samples per client. Although it is quicker to compute, we've already seen that raising the number to just $30$ can significantly improve the resulting group AUC. Looking at the next contenders, it becomes clear that the choice must be made between $30$ and $60$, since when choosing $120$ as the max number of samples per client, the computational times are raised but with a decrease in final group AUC (although not significant). 

Earlier it was argued that "the clear choice seems to be $60$ samples, since it has a statistically higher group AUC when compared to both $30$ ($p_{Bonferroni-adjusted}=3.26e-05$) and $120$ max positive samples although not statistically significant". Setting aside the correctness of the numbers, it does not mean that the choice is clear. Although we can argue that there is a statistically significant difference, it is not that pronounced, and when looking at the computational times, depending on the scenario (number of users and the computational power available), one could argue that using $30$ max positive samples per client would be the wisest choice. It is noticeable that the prediction time is not very much affected, but the biggest tole is observed in the Synthetic Pooling process. Indeed, this is the most challenging part of the data preparation. On the other hand, due to the architecture of the hole system, these synthetic Pools can be pre-processed before any prediction might be needed, since it only looks at past activities. As such, the Synthetic Pooling routine can be scheduled to run in the previous day, and then we only need to account for the prediction time required from the moment we have novel news to score and the moment we are ready to send recommendations. In this case, maitaining the computational power used during research, it would be need around $88$ minutes to predict for $100000$ clients when considering a max number of $60$ positive samples per user and around $87$ minutes to predict for $100000$ clients when considering a max number of $30$ positive samples per user. So, for $100000$ clients, the difference in computational time would be around $1$ minute, not a lot.

\subsection{Variable importance}
Since we are defining a custom feed-forward neural network class and the whole process is done using tensors, with custom tensor loaders for faster iterations, not dataframes, most packages that analyse sample input and result output to inform about variable weights do not work here. Instead, we have opted for a more 'statistical' approach, by measuring the individual AUC scores when using different variables available. These are: the news type, the news category and the news title embeddings. Some relevant variable combinations were tested, namely: the embeddings only (Emb), the type and category without the embeddings (T\&C), the embeddings with category (Emb\&C), the embeddings with type (Emb\&T), and all the variables combined (Emb\&T\&C). 

\begin{figure}[ht]
  \centering
  \includegraphics[width=\linewidth]{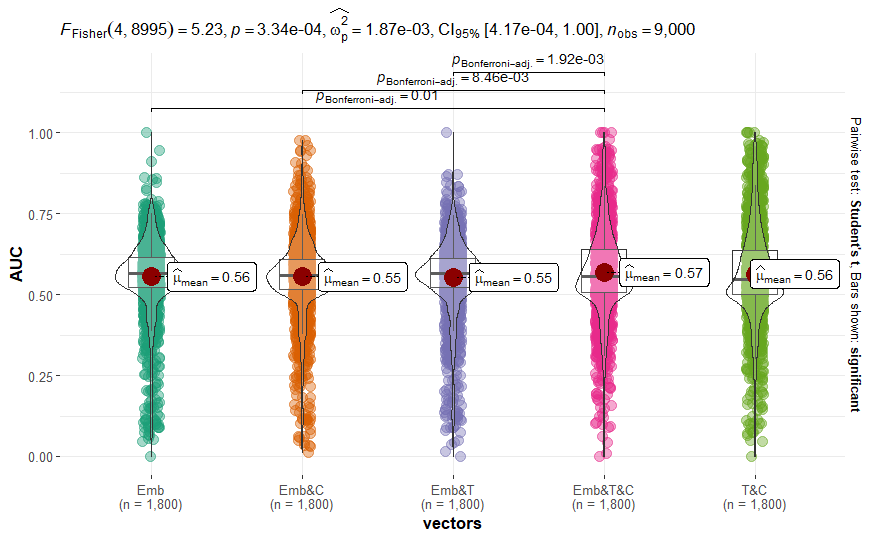}
  \caption{Variable effect on individual AUC}
  \label{fig:vareffect}
\end{figure}

Some eye-opening results can be seen in Figure~\ref{fig:vareffect}, where the distributions are displayed along the corresponding box-plot and violin-plot. Student's \textit{t} was performed for pairwise \textit{post-hoc} testing. Most differences are not statistically significant except for three: Emb and Emb\&T\&C ($p_{Bonferroni-adjusted}=0.01$), Emb\&C and Emb\&T\&C ($p_{Bonferroni-adjusted}=8.46e-03$), and Emb\&T and Emb\&T\&C ($p_{Bonferroni-adjusted}=1.92e-03$). All the differences detected were between using all of the variables combined against using only the title embeddings or these combined with either news category or news type. What is more surprising is that by only using the type and the category information from news, the results come so close to the model using all the variables that there is no statistical significant difference between them. But since using all the variables achieves higher mean AUC, the variance is lower ($0.0176$ vs $0.0209$) and the fact that the title embeddings were already computed for the synthetic negative sampling technique, the choice became clear.

\section{Discussion}
\label{sec:discussion}
Our results can be interpreted from two perspectives, bringing different insights. The first comes directly from the results. The second considers the implications of using a decentralized model.

\subsection{Predictive ability}
In terms of predictive ability, our proposal achieves competitive results when compared to other state of the art news recommenders. 
Another important observation, but tightly connected to the MIND dataset, is that the negative item sampling strategy is key to achieve better performance. We have used a strategy that maximizes the distance of negative items to the observed users' preferences. Changing this strategy to random sampling drastically reduces performance. Moreover, using the negative indicators -- the non-clicked items in the impressions list -- from the dataset has the poorest performance, possibly because these new items already come from a recommended list, so they are likely relevant to the users, just somewhat less relevant than the clicked ones.

\subsection{The road to small user-based networks}

Our method consists of two stages. The first stage is the computation of embeddings from news titles and keywords, which is done centrally. Note that this computation is purely content-based, so it does not require any kind of user data. Then a simple and fast neural network is trained \emph{for each user}, which, again, does not require data to leave the user device(s). Instead, it brings the algorithmic decisions into the user realm. The advantages are three-fold. First, privacy is improved, since no user data is exchanged or stored centrally. Second, user agency is augmented, since users can decide which algorithms to use and how to tune them, at least in theory. Since personal models have lightweight training and inference pipelines, they can run on low-capacity platforms such as mobile phones or web browsers. This may open interesting possibilities in terms of the actual business model involving recommendations. Finally, the decentralized paradigm also brings interesting advantages from the computational perspective, simply because a large share of computation is offloaded to user devices, reducing computational costs and latency, and reducing the dependency on network connectivity.

\subsection{Limitations}
Although our proposal does not strictly require centralized training, we recognize that it is impractical, in most real-world scenarios, to rely exclusively in decentralized computation. The computation of item embeddings, for example, requires access to a very large number of content items, and is too resource-demanding to run in user devices. We note that in the scenario of central computation of embeddings, privacy benefits are maintained, since users are not required to share their history with the provider. It does, however, reduce user agency, since users cannot compute their own item representations. Also regarding the decentralized approach, we note that this is highly facilitated when using a purely content-based method. Collaborative approaches are much harder to decentralize, because they fundamentally rely on personal data from multiple users.

Another thing to keep in mind is that, although this approach can drastically simplify the neural network architectures typically used for these applications, it cannot beat them and it is not perfect. More complex neural architectures should be explored in the future to push the limits of what this training framework can achieve, without compromising its flexibility and easiness to read, implement, maintain and interpret. Indeed, the balance between complexity and performance is a fine line. 

At last, despite the customization power of the proposed approach, people without consolidated reading habits will still receive personalized recommendations based on sparse and seldom past interactions. These interactions can be very old, few and far between. So, even though there is enough information to base the recommendations upon, there is less certainty to the modeled preferences. This might not be ideal if the purpose is to capture the attention of disloyal readers or newcomers. For that, a trending algorithm would probably be more effective than any other alternative, before we could start serving personalized communications.

\section{Conclusion}
\label{sec:conclusions}



In this paper, we propose a decentralized neural News Recommender System approach that explores other less prominent facets of the news recommendation environment. Using the MIND dataset, we show that a thoughtful approach to negative sampling can change the way model architectures are designed and improve model accuracy without the need of adding additional layers of complexity. Our experiments show that while random sampling helps in training models, applying common-sense criteria to negative sampling can yield much better results, and even improve results over the state of the art.   

In addition, by having lightweight individual models for each user, our proposal opens up the possibility to take the recommendation process from the provider's central computing power on to the user's devices, enabling on-device learning. This brings three major benefits: i) reducing the computational cost associated with training pipelines of models on massive amounts of data; ii) the possibility to train profiles without having to transfer private information between users devices and the central computing infrastructure; and iii) the offloading of computation from a centralized infrastructure to user devices in a manageable amount, reducing costs, network dependability and latency.

In future efforts we will seek to improve our approach in several important directions. First, we would like to study the trade-off between model complexity and computation time, taking into account the typical computational constraints of user devices. Second, we will study the integration with a session-based collaborative filtering component, without compromising the benefits of the decentralized approach. Third, a trending algorithm would be a good alternative for clients whose interactions are few and far between. Finally, an online test would be key to measure the performance of these techniques in a real news recommendation scenario.


\bibliography{Bibliography}


\end{document}